\documentclass[12pt]{article} \usepackage{latexsym}
\usepackage{amsfonts} \usepackage{amsmath,cite} \oddsidemargin 0in
\newcommand{\sch}{\mathcal{S}(\mathbb{R}^+)}

\newcommand{\pg}{\psi^G}
\newcommand{\pga}{\psi^G_{appr.}}
\newcommand{\zR}{E_R - i\Gamma_R/2}

\newcommand{\HS}{ {\mathcal{H}}}
\newcommand{\Phx}{ \Phi^\times}
\newcommand{\bk}[2]{ {\langle #1|#2 \rangle}}
\newcommand{\kt}[1]{ {|#1\rangle}}
\newcommand{\br}[1]{ {\langle #1|}}
\newcommand{\kb}[2]{ {| #1\rangle\langle #2|}}
\newcommand{\pbk}[2]{ {\langle {}^{+}#1|#2{}^{+} \rangle}}
\newcommand{\pkt}[1]{ {|#1{}^{+}\rangle}}

\newcommand{\mbk}[2]{ {\langle {}^{-}#1|#2^{-} \rangle}}
\newcommand{\mkt}[1]{ {|#1{}^{-}\rangle}}

\newcommand{\norm}[1]{ {|#1|}}

\begin{document}

\title{Relating the Lorentzian and exponential: Fermi's approximation,
the Fourier transform and causality}

\author{A.~Bohm \\ Physics Department\\ University of
Texas at Austin\\ Austin, TX 78712 \and N.L.~Harshman\\
Department of Physics and Astronomy\\
Rice University\\ Houston, TX 77005 \and H.~Walther\\
Max-Planck Institut f\"ur Quantenoptik und Sektion Physik\\
Universit\"at M\"unchen\\ 85748 Garching, Germany}

\maketitle

\begin{abstract}

\vspace{1cm}

\normalsize

The Fourier transform is often used to connect the
Lorentzian energy 
distribution for resonance scattering to the exponential time
dependence for decaying states.  However, to apply the Fourier transform,
one has to bend the rules of standard quantum mechanics; the
Lorentzian energy distribution must be extended to the full real axis
$-\infty<E<\infty$  instead of being
bounded from below $0\leq E <\infty$ (``Fermi's approximation'').
Then the Fourier transform of 
the extended Lorentzian becomes the exponential, but only for 
times $t\geq 0$, a time asymmetry which is in conflict with the
unitary group 
time evolution of standard quantum mechanics.  
Extending the Fourier transform from distributions to 
generalized vectors, we are led to Gamow kets,
which possess a Lorentzian energy distribution with  $-\infty<E<\infty$
and have exponential time evolution for $t\geq t_0 =0$ only.
This leads to 
probability predictions that do not violate causality.

\end{abstract}

\noindent PACS number(s): 03.65.Db, 03.65.Ca, 32.70.Jz, 06.30.Ft

\section{Introduction}

In this paper we would like to draw a connection between two
theoretical questions concerning the decay of quasistable states.

One question concerns the relation between the two
experimentally-independent signatures of quasistable states, the
Breit-Wigner or Lorentzian lineshape of resonances as measured in the
cross section 
and the exponential decay rate measured for decaying states.
Explicitly, how can the standard relation $\Gamma=\hbar/\tau$ (based
on the Weisskopf-Wigner approximation~\cite{WW}), which
connects the lineshape parameter $\Gamma$ to the lifetime $\tau$, be
justified in a mathematically rigorous way?
Only recently has this relation been experimentally tested to an
accuracy that goes 
beyond the Weisskopf-Wigner approximation~\cite{sodium1,sodium2}.

The other question concerns how the quantum theory of quasistable
states can incorporate the following notion: there can be no
registration of decay products by a detector at times prior to the
preparation of the decaying state.  In other words, we expect that
precursor events should be assigned zero probability in a theory of
quasistable states.  We refer to this common-sense idea as causality
or the
preparation-registration arrow of time~\cite{PRA}.  
Mathematical results within Hilbert space quantum
mechanics challenge the applicability of this notion to quantum
phenomena~\cite{hega} and also challenge the validity of the
exponential law~\cite{Khalfin}.

In the following two sections, we address the above questions in
reverse order and find a connection between them, aided by
the classic example of Fermi's two atom problem~\cite{fermi}.  Then
the next two 
sections present a solution to these questions, defining the
Gamow vectors for the representation of quasistable states.  To make
their definition rigorous, and to incorporate simple answers to the
above questions, entails selecting separate boundary conditions for the space
of states and for the space of observables.  This requires a slight
modification of the Hilbert space axiom of standard 
quantum mechanics.  The consequences, physical and mathematical, of
modifying this axiom so as to incorporate causal decaying state behavior are
discussed in the conclusion.

\section{Probabilities for Precursor Events}

In his classic review of Dirac's quantum theory of radiation~\cite{fermi},
Fermi postulated a problem to test 
whether the theory satisfied what he considered a sensible
requirement, finite-velocity signal propagation (or Einstein causality
as it is occasionally called).  As he posed the problem:
\begin{quote}
Let $A$ and $B$ be two atoms; let us suppose at the time $t=0$, $A$ is
in an excited and $B$ in the normal [ground] state.  After a certain time $A$
emits its energy which may in turn be absorbed by the atom $B$ which
then becomes excited.  Since the light needs a finite time to go from
$A$ to $B$, the excitation of $B$ can take place only after the time
$r/c$, $r$ being the distance between the two atoms.~\cite{fermi}
\end{quote}
Fermi goes on to solve this problem, making the assumption that the
mean life of the excited state of $A$ is short and that the mean life of
the excited state of $B$ is very long.  These facts are used to
justify several simplifications; in particular he uses them to justify
extending the range of integration for the emitted photon frequency
from $[0,\infty)$ to $(-\infty,\infty)$.  Because of this approximation,
several integrals become exact and he achieves his desired result:
there is zero probability that atom $B$ will be excited for $t<r/c$ or
equivalently for $t - r/c <0$.

As we would say today, Fermi achieved his result by analytically
extending the photon energy range.  Though this 
extension of energy appears a quantitatively minor modification,
it has now become clear that precursor events have 
zero probability 
\emph{only} because the integral over the physical
values of energy $0\leq 
E < \infty$ was extended to an integral over the range $-\infty,  < E <
\infty$.
The absence of precursor events is an  
artifact of Fermi's modification of the energy range, not
a consequence of the quantum mechanics of his time.

The root of this problem is not that something ``propagates faster than
light'' to give non-zero transition probabilities for $t<r/c$.  In fact,
if one does not extend the lower bound of the energy range to
$-\infty$, the transition probability is 
non-zero even for $t<0$, i.e., for times before the atom A is excited.  This
result does not indicate a violation of Einstein 
causality so much as a violation of the basic notion of causality
expressed by the preparation-registration arrow of time mentioned in
the introduction.   

This result is not specific to Fermi's two-atom problem, but is the
consequence of a general theorem by Hegerfeldt~\cite{hega}: 
The transition probability between two Hilbert space vectors
is either identically zero for all times (and therefore, there is no
decay at all) or it is different from zero for all time $-\infty <t
<\infty$.  This means that excited states cannot be prepared at a
finite time $t_0 >
-\infty$ and subsequently decay, although this is the typical situation
and the case considered by Fermi.  This Hilbert space theorem has two
consequences: 1.\ non-zero probabilities for precursor events, such as
decay products detected at
$t\leq t_0$ before the state was prepared, a result which violates
causality, and 2.\ non-zero probabilities for detecting decay products
for times $t < t_0 + r/c$ for any finite but arbitrary distance $r$,
or equivalently non-zero probabilities for $r>c(t-t_0)$, which
violates Einstein causality.  This latter consequence was the 
main concern of \cite{hega} though the former consequence more
directly conflicts with the concept of causality in general~\cite{PRA}, 
since it does not also involve space
translation and the constant $c$ of special relativity.  Since
Einstein causality involves the constant $c$ one can only do full
justification to this problem in a relativistic theory using
Poincar\'e transformations, and not just time translations generated by
the Hamiltonian.  We shall briefly discuss this at the end of this
paper and in the Appendix. 

The theorem of \cite{hega} requires that the Hamiltonian be bounded
from below,  
so Fermi circumvented the conditions of this theorem by extending the 
energy values for the spectrum of the Hamiltonian $H$ from $\{E|0\leq
E <\infty\}$ to $\{E| -\infty <
E <\infty\}$ and obtained his result: the probability that
atom B is excited before $t_0=0$, the time at which excited atom $A$ had been
originally created, is zero.
Within standard quantum mechanics
the concept of a time $t_0$ before which the Born probability
is zero is precluded because the (Heisenberg
or Schr\"odinger) equations of motion integrate under the Hilbert space
boundary conditions to the unitary time evolution group, valid for 
$-\infty <t <\infty$~\cite{te}.

Fermi's procedure of extending the energy range can be justified as an
analytic continuation of the energy, but this requires a modification
of one of the axioms of quantum mechanics, the Hilbert space boundary
condition.  In fact, Fermi's approximation points the way to
selecting boundary conditions on the spaces of states and observables
in which the Schr\"odinger and Heisenberg equations do not integrate
to a group, but to a semigroup.

\section{Cross Section and Exponential}

There are two experimental signatures of quasistable states, the
Breit-Wigner or Lorentzian for the energy dependence of the decay
amplitude
\begin{equation}\label{gomega}
g(\omega) = \frac{i}{\omega -
(E_R - i\Gamma/2)}
\end{equation}
and the exponential for the time dependence of the decay amplitude
\begin{equation}\label{expon}
f(t) = e^{-iE_R t}e^{-\Gamma_R t/2}.
\end{equation}

From the point of view of physical observation, the two quantities
$\Gamma$ in (\ref{gomega}) and $\Gamma_R$ (or $\Gamma_R/\hbar$ if the
unit of time is not inverse energy) in (\ref{expon}) are fundamentally 
different quantities and are measured in different ways.
The width $\Gamma$ is measured by the Breit-Wigner
lineshape as, for example, 
in the cross section of a resonance scattering process:
\begin{equation}\label{csbw}
\sigma^{BW}_j(E) \propto \frac{1}{(E-E_R)^2 + (\Gamma/2)^2}.
\end{equation}
The inverse lifetime $\Gamma_R\equiv 1/\tau$, which is the 
initial decay rate (considering only one channel),
can be measured by fitting the counting rate of
decay products to the exponential law:
\begin{equation}\label{excr}
\frac{1}{N}\frac{dN(t)}{dt} = \Gamma_R e^{-\Gamma_R t} \propto e^{-t/\tau},
\end{equation}
where $dN(t_i)$ is the number of decay products registered in the
detector during the time interval $dt$ around $t_i$.

The energy scale of the particular quasistable state usually determines
which of these methods is used to determine the characteristic
lifetime or width parameter\footnote{
There are other experimental methods to determine the lifetime
of a quasistable state, such as using the Primakoff effect for
photoproduction of neutral mesons~\cite{prima} or mapping the resonant
dipole-dipole interaction potential for atomic states~\cite{vleck}.  
We will not consider
these methods and their theoretical connection to the lifetime and
width measurements described here.}.  
For long-lived states, e.g.\ $\tau>10^{-8}\
\mbox{s}$ (or $\hbar/\tau < 10^{-7}$ eV), it is
reasonably easy to directly measure the lifetime by a fit to the exponential
and such states are conventionally called decaying states. 
The range of experimentally accessible lifetimes 
can be extended by exploiting relativistic time dilation
for fast-moving decaying states; for example, a direct lifetime 
measurement has been made of the decay of the 
$\pi^0$~\cite{Atherton}, for which a value of  
$\tau=8.97\pm 0.22 \pm 0.17 \times 10^{-17}$~s was extracted.  This
value, which corresponds to a ratio $\Gamma/E_R \approx 10^{-7}$, is
the limit on lifetime measurements.
For broad resonances, $\Gamma/E_R \approx 10^{-1}\!-\! 10^{-4}$, the
lineshape is easy to measure and
requires an energy resolution of the detection apparatus (and an
energy distribution of the
prepared quasistable states) comparable to $\Gamma/E_R$.
They can be resolved for much
smaller ratios in particular physical systems, such as nuclear
resonances, for which widths have been measured with the M\"ossbauer effect
for at least $\Gamma/E_R < 10^{-15}$ (see table 3.1 of \cite{frau2}
for early data).

Although if one can 
measure the exponential decay rate, the linewidth will be extremely
narrow 
and similarly, although resonances broad enough for their true width to be
measured have very short lifetimes, there have been quastistable
states for which both width and lifetime have been measured and the
lifetime-width relation 
$\Gamma=\Gamma_R(\equiv \hbar/\tau)$ can be experimentally tested.  For
example, using the M\"ossbauer 
effect 40 years ago, the width of the first 
excited state of Fe${}^{57}$ was found to be 4.7$\times
10^{-9}$~eV, which agreed within 10 percent with the direct lifetime
measurement of $\tau=1.4\times 10^{-7}$~s (see section 4.2 of
\cite{frau2}).

A recent precision measurement of $\Gamma = 9.802(22)~\mbox{MHz} = 4.0538(91)
\times 10^{-8}$~eV for the natural linewidth of the
$3p^2P_{3/2}$ excited state of Na (using trapped, ultracold
atoms~\cite{sodium1}) made it possible to compare for the same atomic
state  
the lifetime calculated from $\Gamma$, $\tau=\hbar/\Gamma =
16.237(35)$~ns, with the lifetime measured directly   
$\tau = \hbar/\Gamma_R=16.254(22)$~ns (using beam-gas-laser
spectroscopy~\cite{sodium2}).
The agreement of those two values to an accuracy that exceeds the
accuracy expected of the Weisskopf-Wigner approximation provides
sufficient reason to seek a mathematical theory that justifies the
relation $\Gamma=\Gamma_R=1/\tau$ as an exact equality\footnote{Note
that $\Gamma=\Gamma_R=1/\tau$ 
has never been 
verified in the regime of high energy (relativistic) 
particle physics.  This is of particular interest for 
the Z-boson resonance, where there is much debate
about the parameterization of the
lineshape in terms of mass and width and the relation of the width to
the inverse lifetime (see \cite{nuclphysb} and references 
thereof).}.

Theoretically, the width
$\Gamma$ and inverse lifetime $\Gamma_R$ are often related to each other 
in the following heuristic way: 
the time evolution of a state vector $\phi$ in standard quantum
mechanics with a self-adjoint 
Hamiltonian $H$ is given by
\begin{equation}\label{Ham}
\phi(t) = e^{-iHt}\phi\ \mbox{for} -\infty<t<\infty.
\end{equation}
To obtain from (\ref{Ham}) the exponentials of (\ref{expon}) and (\ref{excr}),
one takes for the time evolution of a decaying state
\begin{subequations}\label{farceinHS}
\begin{equation}\label{dste}
\phi(t) = e^{-i(\zR) t}\phi\ \mbox{for} -\infty<t<\infty
\end{equation}
which would imply that
\begin{equation}\label{complexH}
H\phi = (\zR)\phi.
\end{equation}
\end{subequations}
Ignoring the fact that the statements (\ref{farceinHS})  
cannot hold for a self-adjoint $H$ in standard quantum mechanics, one
can calculate 
the probability amplitude as a function of time
for finding decay products described 
by $\psi$ in
the decaying state $\phi(t)$ (\ref{dste}) as
\begin{equation}\label{decamp}
(\psi,\phi(t)) = (\psi, e^{-iHt}\phi) = f(t) = e^{-iE_Rt}e^{-\Gamma_R
t/2}(\psi,\phi),\ \mbox{for} -\infty<t<\infty.
\end{equation}
Although it has been known for some time
that mathematical theorems of Hilbert space quantum mechanics
forbid amplitudes with the exact exponential time 
dependence~\cite{Khalfin, Fonda}, this heuristic approach is used to
justify the exponential counting rate (\ref{excr}) for decaying
states.  However, more is required if the quasistable state is to have
both the experimental signatures (\ref{gomega}) and (\ref{expon}).

To relate the exponential decay amplitude 
(\ref{expon}) to the Lorentzian, the Fourier transform of the vector
$\phi(t)$ is taken:
\begin{eqnarray}\label{fouvec}
\chi(\omega) &=& \int_{-\infty}^{+\infty}\!\! \!\!dt\,
\phi(t)e^{i\omega t} = \phi \int_{-\infty}^{+\infty}\!\! \!\!dt\ e^{-i(\zR)t}
e^{i\omega t}\\
&=& \frac{i}{\omega-(\zR)} \phi.\nonumber
\end{eqnarray}
Since the vector $\chi(\omega)$ is the function (\ref{gomega}) multiplied
by the vector $\phi$, one argues that the relation between (\ref{gomega})
and (\ref{expon}), and thus the equality $\Gamma=\Gamma_R=1/\tau$, has been 
established theoretically~\cite{WW}.

From the point of view of mathematical rigor, there are
problems with the above heuristic derivation: there is the conflict
between the self-adjoint Hamiltonian and the complex eigenvalue in 
(\ref{farceinHS}) and there is a problem with the boundaries of
integration.
The Fourier transformation is
defined by
\begin{subequations}\label{fou}
\begin{eqnarray}
\mathcal{F}[f(t)]&=& g(\omega)  = \int_{-\infty}^{+\infty}\!\!\!\!dt\;
f(t) e^{i\omega t} \\
\mathcal{F}^{-1}[g(\omega)] &=& f(t) = \frac{1}{2\pi}\int_{-\infty}^{+\infty}
\!\!\!\!d\omega \; 
g(\omega)e^{-i\omega t}
\end{eqnarray}
\end{subequations}
and this is what is usually used in (\ref{fouvec}).
However, the exact mathematical relation between the Lorentzian and
exponential is really given~\cite{Bateman} by
\begin{subequations}\label{exfou}
\begin{equation}\label{geif}
\frac{i}{\omega - (E_R - i\Gamma_R/2)} =
\int_0^{+\infty} \!\!\!\!dt\; e^{-iE_Rt}e^{-\Gamma_R t/2} e^{i\omega t}
\end{equation}
and its inverse is
\begin{equation}\label{igef}
\frac{i}{2\pi}
\int_{-\infty}^{+\infty} \!\!\!\!d\omega \;
\frac{e^{-i\omega t}}{\omega - (E_R - i\Gamma_R/2)} =
\theta(t) e^{-i(\zR)t}
= \left\{ \begin{array}{ll}
			e^{-iE_Rt}e^{-\Gamma_R t/2} & \mbox{for $t\geq
			0$}\\
			0 & \mbox{for $t<0$}\end{array}.
\right.
\end{equation}
\end{subequations}
The ranges of integration
for the variables in the mathematical relation (\ref{exfou}) 
are not the accepted ranges for the
corresponding 
physical quantities in standard quantum mechanics.

The energy, the spectrum of the self-adjoint
Hamiltonian operator, has a lower bound, say $E_0=0$, because the 
scattering energy is a positive
quantity and the stability of matter requires
that the spectrum of the Hamiltonian be bounded from below.  However
the mathematical relation (\ref{igef}) requires the energy range
$-\infty<\omega<+\infty$.
Further, the time variable in quantum 
mechanics extends over $-\infty<t<\infty$ because Hilbert space states
evolve in time by a one parameter unitary group (\ref{Ham}) and this
time evolution is 
reversible.  In the mathematical relation (\ref{geif}),
the values of $t$ range only from $0$ to $\infty$.
So the connection (\ref{exfou}) given by the Fourier transform between
the Lorentzian 
energy distribution and the exponential time dependence of the wave
function requires ranges for time $t$ and energy $\omega$
that are incongruous with what standard quantum theory allows.

Tellingly, this energy range $-\infty<\omega<+\infty$
is exactly what Fermi needed for
his derivation of zero probability for precursor events.  Using the
relation (\ref{exfou}), the probability
amplitude is not given by (\ref{decamp}), but by 
\begin{equation}\label{examp}
(\psi, \phi(t)) = \theta(t)f(t) = \left\{ \begin{array}{ll}
		(\psi, \phi)e^{-iE_Rt}e^{-\Gamma_R t/2} &
			\mbox{for $t\geq 0$}\\
			0 & \mbox{for $t<0$}\end{array}\right.,
\end{equation}
and the notion of time asymmetry, of an initial time $t_0=0$ 
before which there is no decay probability, 
emerges from the mathematics (\ref{exfou}).  

In order to define such an extension of the energy range 
as an analytic continuation of the
energy, we must require certain analyticity properties of the energy
wave functions, and that requires modifying the Hilbert space boundary
condition hypothesis. 
This modification lead to the
definition of the Gamow vector, 
a mathematical representation of unstable states, which we shall use
in place of $\phi$ in (\ref{Ham}).  This Gamow vector has both a
Breit-Wigner energy distribution and an exponential (and asymmetric
as indicated by the $\theta(t)$ in (\ref{igef})) time evolution.

\section{Gamow Vectors}

In this section, we will define the Gamow vectors to be those states
whose energy distributions and time dependence exactly fulfill the
Fourier transform relations (\ref{exfou}).  The decaying Gamow vector has a
Breit-Wigner energy wave function that 
extends over $-\infty<\omega<\infty$ and is related by Fourier
transform to an exponential that starts at finite time $t_0=0$ and
extends only into the future, $t_0<t<\infty$, as required by
causality.
Although these
Gamow vectors cannot be elements of the Hilbert space, they
can be rigorously defined as continuous antilinear functionals on a
dense subspace of the Hilbert space.  They are, like Dirac kets, examples of 
generalized vectors and their use entails the generalization of the
Hilbert space scalar
product.  In parallel with our development of the Gamow vector, we
will show how the same concept of generalized vectors has been used to
give meaning to the Dirac kets.

In standard (time symmetric) quantum theory, a state vector $\phi$ is
an element of 
the Hilbert space, $\phi\in\HS$ (and one also assumes a one-to-one
correspondence between elements of $\HS$ and pure physical states,
c.f.\ (\ref{sqm}) below).
But in Dirac's formalism of quantum mechanics, the state vector is 
expressed in terms of an energy wave function $\phi(E)=\bk{E}{\phi}$
by Dirac's basis vector expansion
\begin{equation}\label{diracE}
\phi = \int_0^\infty \!\!\!\!d\omega\; \kt{\omega}\bk{\omega}{\phi},
\end{equation}
where $\kt{\omega}$ are the Dirac kets which fulfill
\begin{equation}\label{eket}
H\kt{E}=E\kt{E}\ \mbox{for}\ E\in\mbox{spectrum}(H)=\{E|\,0\!\leq E\!<\infty\}.
\end{equation}
The spectrum$(H)$ is the positive real line $\mathbb{R}^+$; it is
bounded from below by $E_0=0$.
 
The scalar product of a vector $\psi\in\HS$ with $\phi\in\HS$ is
\begin{equation}\label{scalar}
\bk{\psi}{\phi}= (\psi,\phi) = \int_0^\infty \!\!\!\!d\omega\;
\bk{\psi}{\omega}\bk{\omega}{\phi}=\int_0^\infty
\!\!\!\!d\omega\;\bar{\psi}(\omega)\phi(\omega).
\end{equation}
In other words, the scalar product is given by a (Lebesgue)
integral over the physical energy values $0\leq
E<\infty$ in the space $L^2(\mathbb{R}^+)$ of energy wave
functions with $\bk{E}{\phi},\bk{E}{\psi}\in L^2(\mathbb{R}^+)$.  The
space $L^2(\mathbb{R}^+)$ is the representation of the physical Hilbert
space $\HS$.

The Dirac ket $\kt{E}$ is not actually an element
of $\HS$.
The Dirac kets can be made mathematically rigorous by defining them
as continuous antilinear functions (often called functionals)
 on a dense nuclear subspace $\Phi$ of the
Hilbert space, $\Phi\subset\HS$. The
Dirac basis vector expansion (\ref{diracE}) is then the Nuclear Spectral
Theorem and holds for all 
$\phi\in\Phi$.
For example, a possible choice for $\Phi$ (and the most common) is such
that $\phi(E)=\bk{E}{\phi}\in\mathcal{S}(\mathbb{R}^+)$, where
$\mathcal{S}(\mathbb{R}^+)$ is
the Schwartz space of 
 rapidly decreasing, infinitely differentiable (smooth) functions restricted to
the positive semiaxis $\mathbb{R}^+$.  Then the ket $\kt{E}$ has
meaning as an element of 
$\Phx$, the space of \emph{continuous antilinear functionals} on $\Phi$, and
its delta function 
energy distribution $\bk{E}{\omega}=\delta(\omega-E)$ is an element of
$(\mathcal{S}(\mathbb{R}^+))^\times$, the space of tempered distributions.
If $\phi$ and $\psi$ are elements of $\Phi$, then their wave functions
are smooth and the integral in the scalar product (\ref{scalar})
is a Riemann
integral.  If $\phi$ and $\psi$ are elements of $\HS$, the integral
in (\ref{scalar}) is a Lebesgue integral because some elements of the
complete Hilbert space $L^2(\mathbb{R}^+)$ will not be smooth.

With the above preparation, we now want to define a vector 
called $\psi^G$ ($G$ for Gamow~\cite{gamow}) whose energy wave function is a
Breit-Wigner distribution,
\begin{equation}\label{approxgamow}
\psi^G = \int \!\!d\omega\; \kt{\omega}\bk{\omega}{\psi^G} =
\int \!\!d\omega\;
\kt{\omega}\left(i\sqrt{\frac{\Gamma_R}{2\pi}}\frac{1}{\omega- (\zR)} 
\right).
\end{equation}
We have two alternatives for
the boundaries of the integration.  If we stick with the rules of standard
quantum mechanics we have to  
choose the boundaries of integration according to (\ref{diracE}) to be
$0\leq \omega<\infty$.  If we want the Fourier
transform to be the exponential, so that $\psi^G$ has an exponential
time evolution then we have to choose the range to cover
$-\infty<\omega<\infty$.

For values of $\Gamma_R/E_R\sim 10^{-1}$ and less, there is not much
numerical difference between these two choices.  For the choice $0\leq
\omega<\infty$, 
which we call the vector $\psi^G_{appr.}$, the scalar product can be
calculated as
\begin{subequations}\label{pgg}
\begin{eqnarray}\label{pga}
(\psi^G_{appr.}, \psi^G_{appr.}) &=& \int_0^\infty \!\!\!\!d\omega\;
\bk{\psi^G_{appr.}}{\omega}\bk{\omega}{\psi^G_{appr.}}
= \int_0^\infty \!\!\!\!d\omega\;
\frac{\Gamma_R}{2\pi}\frac{1}{(\omega-E_R)^2 + 
(\Gamma_R/2)^2}\nonumber\\
&=& \frac{1}{\pi} \int^\infty_{-2E_R/\Gamma_R} dx\frac{1}{x^2 +1} = 1
- \frac{1}{\pi}\left(\frac{\Gamma_R}{2E_R} -
\frac{1}{3}\left(\frac{\Gamma_R}{2E_R}\right)^3 + \cdots\right).
\end{eqnarray}
The vector $\pga$ is a vector in $\HS$, but it is not in the domain of the 
Hamiltonian $H$.

For the choice
$-\infty<\omega<\infty$, which we call $\pg$, we obtain 
\begin{equation}\label{pg}
(\psi^G, \psi^G) 
= \int_{-\infty}^\infty \!\!\!\!d\omega\;
\frac{\Gamma_R}{2\pi}\frac{1}{(\omega-E_R)^2 + 
(\Gamma_R/2)^2} =  1.
\end{equation}
\end{subequations}
Although the numerical difference between (\ref{pga}) and (\ref{pg}) 
is small, the mathematical difference is
enormous.  Although the quantity $(\pg,\pg)$ is a scalar product in
$L^2(\mathbb{R})$, it is no longer a scalar product in 
the Hilbert space $\HS$ represented by $L^2(\mathbb{R}^+)$, since in
(\ref{scalar}) the scalar
product in that space is defined by integration only over the range
$0\leq E<\infty$. 

The difference between the energy distributions is
\begin{subequations}
\begin{equation}\label{edisa}
\norm{\bk{E}{\psi^G_{appr.}}}^2 = \left\{ \begin{array}{ll}
\frac{\Gamma_R}{2\pi}\frac{1}{(E-E_R)^2 + (\Gamma_R/2)^2} & \mbox{for
$0\leq E <\infty$}\\
0 & \mbox{for
$-\infty < E < 0$}\end{array}\right.
\end{equation}
and
\begin{equation}\label{edisb}
\norm{\bk{E}{\psi^G}}^2 = \frac{\Gamma_R}{2\pi}\frac{1}{(E-E_R)^2 +
(\Gamma_R/2)^2}\ \mbox{for}\ -\infty < E < \infty.
\end{equation}
\end{subequations}
For small values of $\Gamma_R/E_R$ the numerical difference between 
(\ref{edisa}) and (\ref{edisb}) for values $E<0$ is negligible.
This is why Fermi may have thought he could neglect this difference and make
his approximation by extending the lower limit of integration to
$-\infty$ without
qualitatively affecting the results.  
But this can be considered a mistake, because there is an
important difference between $\pga$ and $\pg$.   While the
vector $\psi^G_{appr.}$ is a vector in the Hilbert space 
($\bk{E}{\psi^G_{appr.}}\in L^2(\mathbb{R}^+)$), it is \emph{not} an
eigenvector of the Hamiltonian $H$ and it does \emph{not} have exponential
time evolution~\cite{Levitan}.  If one wants an eigenvector of
a self-adjoint $H$ with exponential time evolution, then the
lower limit of the integration must be extended to $-\infty$.  Therefore,
we follow Fermi and extend the boundary of integration in
(\ref{approxgamow}) to define $\pg$:
\begin{subequations}\label{decomg}
\begin{equation}\label{gamow}
\psi^G = \int_{-\infty}^\infty \!\!\!\!d\omega\;
\mkt{\omega}\bk{{}^-\omega}{\psi^G} = 
\int_{-\infty}^\infty \!\!\!\!d\omega\;
\mkt{\omega}\left(i\sqrt{\frac{\Gamma_R}{2\pi}}\frac{1}{\omega- (\zR)} 
\right).
\end{equation}
This vector $\pg$ is now no more an element of the physical Hilbert
space $\HS\Leftrightarrow L^2(\mathbb{R}^+)$, but $\pg$ is a
\emph{generalized} vector 
with ideal Breit-Wigner energy distribution that extends over
$(-\infty, \infty)$~\cite{newref}:
\begin{equation}\label{wf}
\bk{{}^-E}{\pg} = i \sqrt{\frac{\Gamma_R}{2\pi}} \frac{1}{E - (\zR)}\
\mbox{for}\ -\infty<E<\infty.
\end{equation}
\end{subequations}
Unlike the ordinary Dirac kets $\kt{\omega}$ in (\ref{diracE}), we have denoted
the kets in (\ref{decomg}) by $\mkt{\omega}$.  This notation has been
chosen in conformity with the notation for the solutions of the 
Lippmann-Schwinger equation with $-i\epsilon$ in the denominator (the
out-going plane wave solutions).  We will give their exact 
definition after further developing the notion of a generalized vector.

The Gamow vector $\pg$ is generalized in the sense that it is not an element 
of the physical Hilbert space $\HS$, but, like Dirac kets, it has
meaning as a  
continuous antilinear functional $\bk{\psi}{F}=F(\psi)$
 on a dense subspace of the Hilbert space, i.e.\ for
 $\psi\in\Phi\subset\HS$.
Generalizing the scalar product $(\psi,\phi)=\phi(\psi)$ to the bra-ket
$\bk{\psi}{F}=F(\psi)$ introduces a larger class of vectors $F\in\Phx\supset
\HS^\times$ than the continuous antilinear functionals defined by the 
Hilbert space scalar product $(\psi,\phi)=\phi(\psi)$.  This results in a
triplet of spaces, a Rigged Hilbert Space, $\Phi\subset\HS\subset\Phx$; 
the kinds of generalized vectors $\kt{F}$ in the dual space $\Phx$ are 
determined by the choice of $\Phi$.

For the Hilbert space vectors $\phi$, 
the scalar product with all $\psi\in\HS$
is a functional $\phi(\psi)=\bk{\psi}{\phi}=(\psi,\phi)$ on all
$\psi\in\HS$.  It is defined in (\ref{scalar}) (using
Lebesgue integration) and makes mathematical sense for every $\psi\in\HS$.  
However, for the
Dirac ket $\kt{E}\in\Phx$,
\begin{subequations}
\begin{equation}
\kt{E} = \int_0^\infty \!\!\!\!d\omega\;\kt{\omega}\delta(\omega-E),
\end{equation}
the functional  $\bk{\psi}{E}$ on some 
vector $\psi$ makes mathematical sense only if $\psi$ is from a
smaller space $\Phi$.
For such $\psi\in\Phi\subset\HS$, the
wave function $\psi(E)$
\begin{equation}\label{compare}
\overline{\psi}(E) = \bk{\psi}{E} = \int_0^\infty \!\!\!\!d\omega\;
\bk{\psi}{\omega} \delta(\omega-E)
\end{equation}
\end{subequations}
is a smooth function in the Schwartz space, $\bk{E}{\psi}=\overline{\bk
{\psi}{E}}\in\sch$.  
Then, the meaning of Dirac's eigenket equation (\ref{eket}) is that
of a generalized eigenvector equation
\begin{equation}
\bk{H\psi}{E}=\br{\psi}H^\times\kt{E} = E\bk{\psi}{E}\ \mbox{for all}\ 
\psi\in\Phi,
\end{equation}
where $H^\times$ is the unique extension of the operator $H^\dag=H$ to the 
set of functionals $F\in\Phx$.

The vector $\pg$ defined by (\ref{gamow}) is still more generalized
than the Dirac ket $\kt{E}$ and the generalization of the scalar
product, the bra-ket $\bk{\psi^-}{\pg}$, makes
sense only for every $\psi^-$ in a still smaller subspace
$\Phi_+\subset\Phi\subset\HS$.
The space $\Phi_+$ will be chosen such that the energy wave functions
$\mbk{E}{\psi}$ will be smooth Hardy functions in the upper half 
complex energy plane, i.e.\
$\mbk{E}{\psi}\in\HS^2_+\cap\mathcal{S}|_{\mathbb{R}^+}$.
Hardy functions are those smooth functions $\mbk{E}{\psi}$
which can be analytically
continued into the upper complex plane, and $\bk{\psi^-}{E^-}=
\overline{\mbk{E}{\psi}}$ can be continued into the lower half complex plane
and they vanish at the infinite semicircle sufficiently fast~\cite{tit}.
Only for these $\psi^-\in\Phi_+$ does the bra-ket $\bk{\psi^-}{\pg}$ make
sense; i.e.\ we must have $\pg\in\Phx_+$, and then
the value of the functional
$\pg$ at $\psi^-$ is
\begin{equation}\label{gscala}
\bk{\psi^-}{\psi^G} = \frac{i}{2\pi}\int_{-\infty}^\infty\!\!\!\! d\omega\;
\bk{\psi^-}{\omega^-} \frac{\sqrt{2\pi\Gamma_R}}{\omega - (\zR)}
=\frac{1}{2\pi i}
\oint_\mathcal{C} d\omega\; \frac{\sqrt{2\pi\Gamma_R}\bk{\psi^-}{\omega^-}
}{\omega - (\zR)}.
\end{equation}
In the second integral the closed contour $\mathcal{C}$ is from $+\infty$ to
$-\infty$ and then along the infinite semicircle; the equality
follows because $\bk{\psi^-}{\omega^-}$ vanishes on the
infinite semicircle if $\psi^-\in\Phi_+$.

In distinction to the standard Dirac kets $\kt{\omega}\in\Phx$,
the Dirac kets $\mkt{\omega}$ 
used in (\ref{gscala}) (and before in (\ref{decomg})) are 
functionals over the Hardy space $\Phi_+$, i.e.\ $\mkt{\omega}\in\Phx_+$.
This we take as the mathematical definition of the out-going plane wave
solutions of the Lippmann-Schwinger equation. From $\Phi_+\subset\Phi$,
it follows that $\Phx_+\supset\Phx$ and so the set of
$\mkt{\omega}$ is larger than the set of standard Dirac kets; the
functionals $\mbk{\omega}{\psi}$ with $\psi\in\Phi_+$ are defined for any
complex number $\omega$ of the lower half complex plane.  There are also Hardy 
spaces $\Phi_-$ for which the kets $\pkt{\omega}$ are defined for any
complex number $\omega$ of the upper half complex plane; the role of these
spaces will be discussed in the final section.

Evaluating formula (\ref{gscala}) using the
Cauchy formula~\cite{tit}, we obtain
\begin{equation}\label{gscalb}
\bk{\psi^-}{\pg}=\frac{i}{2\pi}\int_{-\infty}^\infty\!\!\!\! d\omega\;
\bk{\psi^-}{\omega^-} \frac{\sqrt{2\pi\Gamma_R}}{\omega - (\zR)} =
 \bk{\psi^-}{\zR 
{}^-}\sqrt{2\pi\Gamma_R}.
\end{equation}
This means that
$\pg=\sqrt{2\pi\Gamma_R}\mkt{\zR}\in\Phx_+$ is something
like a Dirac ket extended to  
the complex value $\zR$.  Also, comparing this with equation (\ref{compare})
 shows that the Breit-Wigner distribution
$i/2\pi(E-(\zR))^{-1}$ is something like a Dirac $\delta$-function when
applied to the well-behaved Hardy functions, i.e.\ just as integrating over
$\delta(\omega-E)$ maps every function 
$\bk{\psi}{\omega}\in\mathcal{S}$ to its value at $E$, integrating over
$i/2\pi(\omega-(\zR))^{-1}$ maps every function
$\bk{\psi^-}{\omega^-}\in\HS_-^2\cap\mathcal{S}|_{\mathbb{R}^+}$
to its value at $\zR$ with $\Gamma_R\geq 0$.

The properties of  well-behaved Hardy class vectors ensure that if
$\psi^-\in\Phi_+$, then also $H\psi^-\in\Phi_+$.  Then replacing
$\psi^-$ in (\ref{gscalb}) with $H\psi^-$ (or equivalently
replacing $\bk{\psi^-}{\omega^-}$ with $\bk{H\psi^-}{\omega^-}=
\br{\psi^-}H^\times\mkt{\omega}=\omega\bk{\psi^-}{\omega^-}$)
one proves
\begin{equation}\label{evecta}
\bk{H\psi^-}{\pg}=\br{\psi^-}H^\times\kt{\psi^G} = (\zR)\bk{\psi^-}{\psi^G}\
\mbox{for all 
$\psi^-\in\Phi_+$}.
\end{equation}
This proves that the Gamow vector $\psi^G$ is a generalized eigenvector of the
self-adjoint and semi-bounded Hamiltonian $H$ with a complex
eigenvalue (note that this cannot be proved for $\pga$; it is not an
eigenket of $H$).
Omitting the arbitrary
$\psi^-\in\Phi_+$ in (\ref{evecta}), the eigenvector property is
written in the Dirac notation as
\begin{equation}
H^\times\kt{\psi^G} \equiv H^\times
\mkt{\zR}\sqrt{2\pi\Gamma_R}=(\zR)\kt{\psi^G}.
\end{equation}
Such eigenvectors (eigenkets) of a self-adjoint Hamiltonian with
complex eigenvalue do 
not exist in the Hilbert space or even in $\Phx$, the dual to the
Schwartz space, but they are in $\Phx_+\supset\Phi_+$,
the dual to the well-behaved
Hardy class space $\Phi_+$ of the upper half complex plane.

\section{Exponential Decay Law from Breit-Wigner Distribution}

As shown above, the Gamow vectors are eigenvectors of a self-adjoint 
Hamiltonian with complex eigenvalues, thus justifying the 
heuristic equation (\ref{complexH}).  However this is not enough
to justify the exponential time evolution (\ref{dste}); for that we
return to the Cauchy formula.

Note that if
$\bk{\psi^-}{\omega^-}\in\HS_-^2\cap\mathcal{S}|_{\mathbb{R}^+}$, then  
$\exp(-i\omega t)\bk{\psi^-}{\omega^-}$ will also be in
$\HS_-^2\cap\mathcal{S}|_{\mathbb{R}^+}$, 
though \emph{only} for $t\geq 0$.  For $t<0$, the exponential factor blows up
as $\omega$ approaches the infinite semicircle.  With this fact, the
time evolution of the Gamow vector 
is calculated for all $\psi^-\in\Phi_+$ as follows:
\begin{eqnarray}\label{evol}
\bk{e^{iHt}\psi^-}{\zR} &=& \br{\psi^-}e^{-iHt}\mkt{\zR}\nonumber\\ 
&=&
\frac{1}{2\pi} \int_{-\infty}^{+\infty}\!\!\!\!d\omega\;
\br{\psi^-}e^{-iHt}\mkt{\omega} 
\frac{1}{\omega-(\zR)}\nonumber\\
&=& \frac{1}{2\pi} \int_{-\infty}^{+\infty}\!\!\!\!d\omega\;
e^{-i\omega t}
\bk{\psi^-}{\omega{}^-} 
\frac{1}{\omega-(\zR)}\nonumber\\
&=& e^{-iE_Rt}e^{-\Gamma_Rt/2}\bk{\psi^-}{\zR{}^-}\\
&&\mbox{for
all}\ \psi^-\in\Phi_+\ \mbox{and for}\ t\geq 0\ \mbox{only}.\nonumber
\end{eqnarray}
The last step in equality (\ref{evol}) 
is the generalization of the formula for the Fourier 
transform of the Lorentzian
(\ref{igef}), which one observes if one multiplies
the Lorentzian in (\ref{igef}) 
by a Hardy class function of the lower complex energy
plane, i.e.\ by a function
$\bk{\psi^-}{\omega^-}\in\HS^2_-\cap\mathcal{S}|_{\mathbb{R}^+}$:
\begin{eqnarray}\label{genigef}
\frac{i}{2\pi} \int_{-\infty}^\infty\!\!\!\! d\omega\;
 e^{-i \omega t}
\frac{\bk{\psi^-}{\omega^-}}{\omega - (\zR)} &=& \theta(t)e^{-i(\zR)t}
\bk{\psi^-}{\zR{}^-} \\
&& \mbox{for any}\
\bk{\psi^-}{\omega^-}\in\HS^2_-\cap\mathcal{S}|_{\mathbb{R}^+}.\nonumber
\end{eqnarray}

The mathematical result (\ref{evol}) for the Gamow ket
$\psi^G=\kt{\zR{}^-} \sqrt{2\pi\Gamma_R}$ can also be written in the
Dirac notation if one omits the arbitrary $\psi^-\in\Phi_+$:
\begin{eqnarray}\label{elaw}
e^{-iH^\times t}\mkt{\zR} &=& 
\int_{-\infty}^{+\infty}\!\!\!\!d\omega\;
\frac{i}{2\pi}\frac{e^{-i\omega t}}{\omega-(\zR)}\mkt{\omega}\\
&=& e^{-iE_Rt}e^{-\Gamma_Rt/2} \mkt{\zR}\ \ \ \mbox{for $t\geq 0$ only}.
\nonumber
\end{eqnarray}
The equality in the second line of (\ref{elaw}) is the vector
analog  of the
Fourier transformation formula (\ref{igef}), and (\ref{genigef}) is the
wave function analog of (\ref{igef}).
It expresses the exponential law for the time evolution of the vector
$\psi^G(t)$ with idealized Breit-Wigner energy distribution.

Thus the Gamow vector is needed to accomplish the task posed:
derive the exponential law from the Lorentzian energy distribution.
The well-known Fourier transformation formula (\ref{exfou}) suggests
that we should follow Fermi's ``mistake'' and extend the energy to $-\infty$.
As can be expected from the $\theta(t)$ in
(\ref{exfou}), time asymmetry about the time $t=0$ (which can be any time
$t_0\neq -\infty$) results.

Such a privileged moment of time $t=0$ cannot exist in standard quantum 
mechanics.  The time evolution in the Hilbert space is necessarily given 
by the reversible, unitary group of operators
$U^\dag(t)=\exp(-iHt)$~\cite{te}; with every $U^\dag(t)$ there exists also an
inverse
$(U^\dag(t))^{-1}=U^\dag(-t)$.
However, in the space $\Phx_+$, the time evolution operator
$U^\times(t)\equiv\exp(-iH^\times t)
$ (\ref{elaw}) is the uniquely defined extension of the unitary 
operator $U^\dag(t)$ for $t\geq 0$ \emph{only}; $(U^\times(t))^{-1}$ is not
defined on $\Phx_+$ and $U^\times(t)$ is a semigroup.  The 
mathematical necessity of this time asymmetry and special time $t=0$
has a physical consequence: the semigroup time 
transformation properties of the Gamow vector (\ref{elaw}) 
have the notion of time's arrow (or irreversibility) and a privileged
time $t_0$ built in.

\section{Conclusion}

In summary, the derivation in introductory textbooks which relates the
Breit-Wigner energy distribution to the exponential time evolution, and 
which makes no sense in standard mathematical theory
of quantum mechanics in the Hilbert space, can be given a mathematical meaning.
To do so leads to the Gamow vectors and their irreversible 
semigroup time evolution.  How one has to construct the Gamow vector is
indicated by the mathematical relations (\ref{exfou}) for the
Fourier transform between Lorentzian and exponential.  The relation
(\ref{exfou}) 
already contains
the first appearance of time
asymmetry and forebodes the inadequacy of the unitary time
evolution of standard quantum mechanics.  The time evolution of the
Gamow vectors (\ref{elaw}) can be viewed as just the generalized
vector version of 
the relation (\ref{exfou}) between generalized functions.

To incorporate generalized vectors requires a partial revision of the
mathematical theory of quantum physics.  The Hilbert space is a vector
space with topology (i.e.~the notion of convergence for infinite
sequences of vectors)  
given by norm convergence.  While the 
appropriateness of a linear space of states is experimentally verified
by the glorious success of the superposition principle, the appropriateness
of the norm topology cannot be directly tested due to the
limited resolution of experimental equipment and the infinite number of
measurements required to make a topological distinction.  
It is the algebraic properties of a linear scalar product space, and not
the topological properties, 
to which most physicists refer when discussing the Hilbert space.  The
Hilbert space results when
this algebraic structure is completed with the norm topology, and then
the mathematical theory used in \cite{hega,Khalfin,te} is obtained.

The first revision of this mathematical theory of
quantum mechanics was made so that elements like the Dirac ket could be
incorporated; it is now well-accepted and uses the Rigged Hilbert space
\begin{equation}\label{lamerhs}
\Phi\subset\HS\subset\Phx
\end{equation}
where $\Phi$ is usually chosen to be the Schwartz space (which
has a stronger topology than $\HS$) and then $\Phx$, the 
space of continuous antilinear functionals on $\Phi$, is the
space of tempered distributions~\cite{RHS}.

The scalar product $\norm{(\psi,\phi)}^2$ in the
Hilbert space is interpreted as the probability to detect the
observable $\kb{\psi}{\psi}$ in the state
$\phi$.  For the generalized ``scalar product'' with a Dirac ket, 
the physical interpretation of
$\norm{\bk{E}{\psi}}^2$ is as the probability density 
in energy using an apparatus described by $\psi$.  Since the
macroscopic apparatus will have a smooth energy
resolution, the
amplitude $\bk{E}{\psi}$ is best described by a smooth
function, $\bk{E}{\psi}\in\mathcal{S}(\mathbb{R}^+)$ or equivalently
$\psi\in\Phi$.  Thus the mathematic necessity of using the space
$\Phi$ to define the Dirac kets has a physical justification.

The second step of this revision is to incorporate the Gamow vector.
In order to include generalized vectors with 
complex (and also negative) energies, the energy
wave functions $\mbk{E}{\psi}$ cannot be just smooth functions but
must also be analytically continuable.  Instead of using the Hilbert space 
axiom
\begin{equation}\label{sqm}
\{\mbox{space of prepared states}\} = \{\mbox{space of detected observables}\}
=\HS
\end{equation}
or the slightly more general revision
\begin{equation}\label{ssqm}
\{\mbox{space of prepared states}\} = \{\mbox{space of detected observables}\}
=\Phi\subset\HS,
\end{equation}
we distinguish mathematically
between states and observables~\cite{feyn} and make the new
hypothesis~\cite{rhsrev}:
\begin{eqnarray}\label{ourbc}
\mbox{The prepared states are described by:}\
&\{\phi^+\}&=\Phi_-\subset\HS\subset\Phx_-\\
\mbox{and the registered observables by:}\
&\{\psi^-\}&=\Phi_+\subset\HS\subset\Phx_+.\nonumber
\end{eqnarray}
Here we use a pair of
Rigged Hilbert spaces of Hardy type,
where the energy wave functions of the vectors $\phi^+\in\Phi_-$ and 
$\psi^-\in\Phi_+$ are well-behaved Hardy functions in the lower and upper half
complex planes, respectively~\cite{gad}.  
The Gamow vectors, together with the out-plane wave solutions of the
Lippmann-Schwinger equations, are elements of the space $\Phx_+$.

The states prepared by the preparation apparatus are, by hypothesis
(\ref{ourbc}), not generalized vectors like the Gamow vectors
(\ref{decomg}); the states $\phi^+\in\Phi_-\subset\HS$ are represented by the
Dirac basis vector expansion (\ref{diracE}) with wave functions
$\phi^+(\omega) = \pbk{\omega}{\phi}$ that are well-behaved Hardy
functions in the lower half plane.  From (\ref{ourbc}) one can
prove (see the reference in citation \cite{tit}) that an alternate
``complex basis 
vector'' expansion 
holds for every $\phi^+\in\Phi_-$:
\begin{equation}\label{cbve}
\phi^+ = \sum^N_{i=1} \kt{\pg_i}\bk{\pg_i}{\phi^+} +
\int_{-\infty}^{+\infty}
\!\!\! d\omega\, \pkt{\omega}\,b(\omega).
\end{equation}
Here, the $\pg_i$ represent $N$ Gamow vectors with eigenvalues
$z_{R_i} = E_{R_i} - i\Gamma_i/2$ which are associated to $N$ first order
resonance poles at the pole positions $z_{R_i}$ of the analytically
continued S-matrix.  However, in addition to the sum over the
resonance states $\pg_i$, there appears an integral over the
continuous basis vectors with weight function $b(\omega)$.  Whereas each
Gamow vector in (\ref{cbve}) corresponds to a Breit-Wigner resonance
amplitude, the integral corresponds to a slowly varying background in
the scattering amplitude.

Considering just one resonance ($N=1$), then
the first term of (\ref{cbve}) represents the state of the resonance
with its characteristic exponential time behavior.  However, there is
always also a background term whose time evolution is not exponential
and whose energy-dependence $b(\omega)$ depends on the way in which
the state was prepared.  This background term is a theoretical
necessity but in a particular experiment may be unimportant. The
influence of this background can
explain observed deviations from the exponential decay
law~\cite{devexp}, which had been derived mathematically a long time
ago~\cite{Khalfin} for Hilbert space vectors, and thus also for
$\phi_+\in\Phi_-\subset\HS$.  But the Gamow vectors are elements of
$\Phx_+$ and are not in $\HS$, and therefore their time evolution can
follow the time-honored exponential law.

In the spaces $\Phi_-$ and $\Phx_-$ and in the space $\Phi_+$ and
$\Phx_+$ (but not in $\HS$) only semigroup time evolution is defined.
The semigroup time evolution  is the crucial difference
between standard time symmetric quantum mechanics in the Hilbert
space or in a triplet such as (\ref{lamerhs}) and a quantum theory that
includes time asymmetry.  By changing the 
boundary conditions, i.e.\ the space of allowed solutions of the dynamical
equations (Schr\"odinger or Heisenberg) from (\ref{sqm} or \ref{ssqm}) to 
(\ref{ourbc}), we obtain semigroup evolution.  Time asymmetric boundary 
conditions of time symmetric dynamical (differential) equations are nothing
new in physics.  The cosmological arrow of time and the radiation arrow of
time are consequences of such a theory.  The idea of a fundamental arrow of
time in the laws of quantum mechanics is also nothing new, c.f.~\cite{GM}
where
it had to be affixed artificially on top of the time symmetric solutions
because the Hilbert space boundary condition for the quantum mechanical Cauchy 
problem excluded time asymmetry.
With the hypothesis (\ref{ourbc}), asymmetric time evolution is a
consequence of the mathematical property of the Hardy space
functions~\cite{PRA} (specifically the Paley-Wiener theorem for their
Fourier transform~\cite{tit}) in the same way as reversible time
evolution is a mathematical consequence of the properties of Hilbert space 
functions~\cite{te}.

The Gamow vectors $\pg$ describe
resonance states (without background).  They are the only 
mathematical entity which can
exactly combine
the properties of Lorentzian 
energy distribution and exponential time evolution.
The relation between the
lifetime $\tau$ (from the counting rate (\ref{excr})) and the Lorentzian width
$\Gamma$ (from the cross section (\ref{csbw})) is precisely and exactly
$\tau=\hbar/\Gamma$ and this lifetime-width relation holds
exactly only for the Gamow vectors.

The Gamow vectors also predict causal probabilities.  As a consequence
of their semigroup time
evolution (\ref{evol},\ref{elaw}) the concept of a finite
time $t_0>=\infty$ (represented by the semigroup time $t=0$) is
introduced for a quantum
system described by $\pg$, and before that time the
quantity 
$\bk{e^{iHt}\psi^-}{\pg} = \bk{\psi^-}{e^{-iHt}\pg} = \bk{\psi^-}{\pg(t)}$
does not exist.
Extending the usual probability interpretation of
$\norm{(\psi,\phi)}^2$ to the Gamow vector, the
generalized scalar product 
$\norm{\bk{\psi^-}{\pg(t)}}^2$ 
describes the probability to
detect the decaying quantum state $\pg=\mkt{\zR}\sqrt{2\pi\Gamma_R}$ 
at the time $t$ with an apparatus
described by $\psi^-$.  Lets put this into a specific physical context
for the purpose of example by considering the process where a $K^0$ is
created by $\pi^- p\rightarrow K^0\Lambda$ and then decays via the
channel $K^0\rightarrow\pi^+\pi^-$.  The Gamow vector $\pg$ represents the
decaying state $K^0$ and $\psi^-$ represents the the decay products 
$\pi^+\pi^-$ which the detector
registers.  Since the decay
products cannot be detected before the decaying state $\pg$ has been
created (or prepared) at an arbitrary but finite time $t_0=0$, the
probability to detect the decay product $\psi^-$,
$\norm{\bk{\psi^-}{\pg(t)}}^2=\norm{\bk{e^{iHt}\psi^-}{\pg}}^2$, 
makes
physical sense only for $t\geq 0$.  This means that for the decaying
kaon system, we
should expect a non-zero counting rate for the decay products
$\pi^+\pi^-$ only for times \emph{after} the $K^0$ has been created in
the reaction $\pi^- p\rightarrow K^0\Lambda$ and leaves the proton
target.
Therefore the probability for this decay
$\norm{\bk{\psi^-}{\pg(t)}}^2$ must be different from
zero only 
for $t>t_0=0$ and this is precisely the
time asymmetry predicted by (\ref{evol}).
Thus the semigroup time $t=0$ is interpreted as the time $t_0$
at which the creation of
the decaying state is completed and the registration of the decay
products can begin.  For the $K^0$ system this time is very accurately
measurable because the $K^0$ is created by the strong interaction with
a time scale of $10^{-23}$s and it decays weakly with a time scale of
$10^{-10}$s. 

The result (\ref{evol}) predicts that the probability
$\norm{\bk{\psi^-}{\pg(t)}}^2=
\norm{\br{\psi^-}U^\times(t)\kt{\pg(t)}}^2$ to detect decay
products $\psi^-$ from a state $\pg$ which has been
prepared at $t_0$ is zero for $t<t_0$.  
Therefore the probability of precursor events is zero, in contrast to
the Hilbert space result obtained in \cite{hega} which does predict
precursor events.  This feature is what we mean by causality.

This prediction applies 
when the detector  $\kb{\psi^-}{\psi^-}$ is 
placed right at the position of the decaying particle $\psi^G$ and
is in the rest frame of the decaying state.  If the decaying state and
detector are displaced or have relative motion, then time translation
alone cannot describe the situation and Poincar\'e transformations must
be considered.  
To represent decaying states that undergo relativistic
transformations, relativistic Gamow vectors were introduced~\cite{rgv}.
Unlike the relativistic stable particle states which furnish a unitary
(Wigner) representation of the Poincar\'e group~\cite{wigner},
relativistic Gamow vectors furnish only a semigroup representation of
the Poincar\'e transformations into the forward light cone
\begin{equation}\label{flc}
x^2 = t^2 -\mathbf{x}^2 \geq 0,\ t\geq 0,
\end{equation}
(c.f.~Appendix and references thereof).
One therefore predicts in this relativistic theory space-time
translated probability amplitudes
\begin{equation}
\br{\psi^-}U^\times(1,(t,\mathbf{x}))\kt{\pg} =
\br{\psi^-}e^{-i(H^\times t - \mathbf{P}\cdot\mathbf{x})}\kt{\pg} 
\end{equation}
\emph{only} for those space-time translations $(t,\mathbf{x})$ which fulfill
(\ref{flc}).

This is the relativistic analogue of the time asymmetry
$t\geq 0$ in (\ref{evol}).  Therefore, decay events of the space-time
translated 
state $U^\times(1,(t,\mathbf{x}))\kt{\pg}$ are only predicted to occur
for $t\geq 0$ and $t^2\geq \mathbf{x}^2 \equiv r^2/c^2$ or for $r/t\leq
c$.  This means the probability for decay events
cannot propagate faster than the speed of light and Einstein causality
is obeyed by these semigroup representations of the Poincar\'e 
transformations.  This confinement to semigroup transformations
into the forward light cone 
follows from the Hardy space hypothesis 
(\ref{ourbc}).  
Unitary representations of the Poincar\'e group in the Hilbert space
are not restricted to the forward light cone (\ref{flc}) and therefore
do not fulfill Einstein causality.  This can be deduced from
the theorem in reference \cite{hega} which is based on the Hilbert
space axiom (\ref{sqm}).  The
Hardy space hypothesis predicts probabilities only for 
semigroup time evolution (for non-relativistic) or Poincar\'e 
semigroup evolution in the forward light cone (for relativistic) and
as a consequence both the causality conditions ``no
registration of an observable in a state before that state has been
prepared'' and the Einstein causality condition ``no propagation of
probabilities with a speed faster than light'' are fulfilled.

Fermi's fortuitous (but at the time unjustified) approximation, which was
made so that causality was maintained,
pointed the way to the Gamow vectors, and the mathematical relation of the
Fourier transform (\ref{exfou}) foretold the time asymmetry.

\section*{Appendix: Poincar\'e Transformations of \\ Relativistic
Resonance States}
\setcounter{equation}{0}
\renewcommand{\theequation}{A\arabic{equation}}

To address Einstein causality for the representation of resonances and
decaying states, we 
must consider Poincar\'e transformations of a relativistic Gamow
vector.  We will therefore briefly record some definitions and
results.  Since this subject exceeds the scope of the present paper we shall
just review the results here and refer to \cite{rgv} for details.  

The non-relativistic Gamow vector was in some sense a
generalization of the Dirac ket to complex energies and so we begin by
summarizing the representations of
relativistic stable states.
Stable particles are described by irreducible
unitary representation spaces of the Poincar\'e group characterized by the
invariant mass squared $m^2$ and by spin
$j$~\cite{wigner}.  The basis vectors
$\kt{[j,m^2]\mathbf{p}j_3}$ of an
irreducible representation space are usually labelled by
a component of the spin $j_3$ and
the spatial components $\mathbf{p}$ of the 4-momentum $p=(p_0,
\mathbf{p})$.  Instead of $\mathbf{p}$ one could 
equivalently use 
the spatial components $\mathbf{\hat{p}}$ of the 4-velocity
$\hat{p}=p/m=(\gamma, \mathbf{\hat{p}})= (\gamma, 
\gamma\mathbf{v})$ and then use the 4-velocity eigenkets 
$\kt{[j,m^2]\mathbf{\hat{p}}j_3}$ as basis vectors.

The relativistic Gamow vector $\mkt{[j,\mathsf{s_R}]\mathbf{\hat{p}}j_3}$
is defined from the resonance pole $\mathsf{s}_R=(M - i\Gamma/2)^2$
of the S-matrix for a resonance  
scattering process $1+2\rightarrow R \rightarrow 3+4$ by
\begin{equation}\label{relg}
\mkt{[j,\mathsf{s_R}]\mathbf{\hat{p}}j_3} = \frac{1}{2\pi}
\int^{+\infty}_{-\infty_{II}} 
\frac{d\mathsf{s}}{\mathsf{s} - \mathsf{s}_R}
\mkt{[j,\mathsf{s}]\mathbf{\hat{p}}j_3}. 
\end{equation}
Here $\mathsf{s} = (p_1 +
p_2)^2$ is the invariant mass squared and the integration extends
along the lower edge of the real $\mathsf{s}$-axis in the second sheet
(the sheet that contains the pole at 
$\mathsf{s}_R$).  The definition
(\ref{relg}) is the relativistic analogue of (\ref{gamow}).
The kets $\mkt{[j,\mathsf{s}]\mathbf{\hat{p}}j_3}$ are the
out-plane wave solutions of the Lippmann-Schwinger equation  
and are functionals over a
Hardy space $\Phi_+$ (which is slightly different than the space
$\Phi_+$ chosen for the non-relativistic case but has the same
analyticity properties), whereas the usual Wigner basis kets for
unitary representations are functionals over the Schwartz space.

The transformation properties of the relativistic Gamow vectors under
Poincar\'e transformations $(\Lambda,x)$ (where $x=(t,\mathbf{x})$)
are given by 
\begin{subequations}
\begin{equation}\label{gamwiggy}
U^\times(\Lambda, x)\mkt{[j,\mathsf{s_R}]\mathbf{\hat{p}}j_3}= 
e^{-i\gamma\sqrt{\mathsf{s}_R}(t - \mathbf{x}\cdot\mathbf{v})}\sum_{j'_3}
D^j_{j'_3
j_3}(W(\Lambda^{-1},\hat{p}))
\mkt{[j,\mathsf{s_R}]\mathbf{\Lambda^{-1}\hat{p}}j'_3}
\end{equation}
only for
\begin{equation}\label{flca}
t\geq 0\ \mbox{and}\ t^2 \geq
\mathbf{x}^2.
\end{equation}
\end{subequations}
The matrix $D^j_{j'_3 j_3}$ is the $(2j +1)$-dimensional
representation of the Wigner rotation
$W(\Lambda^{-1},\hat{p}) =
L^{-1}(\Lambda\hat{p})\Lambda L(\hat{p})$ and $L(\hat{p})$ is the
standard boost, which depends only on $\hat{p}$ and not the momentum
$p=\sqrt{\mathsf{s}_R}\hat{p}$.  It is this property that allows
construction of the representation $[j,\mathsf{s}_R]$ by analytic
continuation of the Lippmann-Schwinger kets to the Gamow kets,
\begin{equation}
\mkt{[j,\mathsf{s}]\mathbf{\hat{p}}j_3}
\rightarrow\mkt{[j,\mathsf{s_R}]\mathbf{\hat{p}}j_3}
\end{equation}
in such a way that $\hat{p}$ remains unaffected and always real.
These representations $[j,\mathsf{s}_R]$ are the ``minimally complex''
representations.

In the limit as the complex invariant mass squared $\mathsf{s}_R$ becomes real,
$\mathsf{s}_R=(M - i\Gamma/2)^2\rightarrow m^2$, the transformation
formula (\ref{gamwiggy}) looks exactly like the well-known formula for
Wigner's unitary 
representations $[j,m^2]$:
\begin{equation}\label{wiggy}
U^\times(\Lambda,x)\mkt{[j,\mathsf{s}]\mathbf{\hat{p}}j_3} = e^{-ip\cdot x}
\sum_{j'_3} D^j_{j'_3 j_3}(W(\Lambda^{-1},\hat{p}))
\mkt{[j,\mathsf{s}]\mathbf{\Lambda^{-1}\hat{p}}j'_3},
\end{equation}
where $\exp(-ip\cdot x) = \exp(-i\gamma m(t -
\mathbf{v}\cdot\mathbf{x}))$.  However, the important difference
between (\ref{wiggy}) and Wigner's transformation formula for unitary
representations is that Wigner's transformation formula holds for the
whole Poincar\'e 
group
\begin{equation}
\mathcal{P} = \{(\Lambda, x): \Lambda\in\mathrm{S0}(1,3),
x\in\mathbb{R}_4\}
\end{equation}
whereas the transformation formulas (\ref{wiggy}) for the
Lippmann-Schwinger kets and (\ref{gamwiggy}) for the Gamow kets hold
only for the 
orthochronous Lorentz
transformations and space-time translations into the forward light
cone:
\begin{equation}
\mathcal{P}_+ = \{(\Lambda, x): \Lambda\in\mathrm{S0}(1,3),
x\in\mathbb{R}_4| x^2 = t^2 -\mathbf{x}^2 \geq 0, t\geq 0\}.
\end{equation}
The formulas (\ref{gamwiggy},\ref{flca}) are the relativistic
generalizations  of the
transformation formula (\ref{elaw}) so as to include all allowed space-time
translations, which, as for the time translations in (\ref{elaw}), form
only a semigroup 
$\mathcal{P}_+$.

Specializing (\ref{gamwiggy}) to space-time translations $(1,x)$,
we obtain from 
(\ref{gamwiggy})
the space-time translated probability amplitude in analogy to (\ref{evol}):
\begin{eqnarray}
\bk{\psi^-(x)}{\mathbf{[j,\mathsf{s_R}]\hat{p}}j_3 {}^-} &=&
\bk{U(1,x)\psi^-}{[j,\mathsf{s_R}]\mathbf{\hat{p}}j_3{}^-}\nonumber\\
&=& e^{-i\gamma(M - i\Gamma/2)(t - \mathbf{x}\cdot\mathbf{v})}
\bk{\psi^-}{\mathbf{[j,\mathsf{s_R}]\hat{p}}j_3 {}^-}\ \mbox{for all}\ \psi^-\in\Phi+
\end{eqnarray}
but only for $t\geq 0$ and $t^2\geq \mathbf{x}^2$ because the Hardy
space $\Phi_+$ does not contain vectors $U(1,x)\kt{\psi^-}$ for which
(\ref{flca}) is not fulfilled.
Introducing the speed of light, the
condition for the predicted probability (\ref{flca}) becomes $t\geq 0$
and $c^2t^2 \geq 
\mathbf{x}^2$ or $|\mathbf{x}|/t \leq c$.  This means the probability
for decay events is only predicted for the forward light cone and 
cannot propagate faster than the speed of light.  Thus Einstein causality
is not violated.

\section*{Acknowledgements}

This paper stems from discussions following the lecture of one of us (A.B.) at
the Max Planck Institut f\"ur Quantenoptik and we would like to express 
our gratitude to the Humboldt Foundation whose award made this visit possible.
We are grateful to W.~Drechsler for discussions and drawing our 
attention to \cite{feyn}.  Discussions with R.~Hulet and M.~Mithaiwala
provided insight and information and Duane
Dicus and G.~C.~Hegerfeldt read the manuscript 
at various stages and 
made many valuable suggestions.  We would also like to thank the Welch
Foundation for additional financial support.

\end{document}